\documentclass{crc-book}

\def\ii#1{{}}

\def\a{\alpha}
\def\b{\beta}

\def\pa{\partial}

\def\eeq{\end{equation}}
\def\beq{\begin{equation}}

\def\eqref#1{(\ref{#1})}

\def\EOR{ \hfill $\odot$}

\begin{document}

\renewcommand{\evenhead}{G. Gaeta}
\renewcommand{\oddhead}{Geometry of Normal Forms}

\thispagestyle{empty}

\Name{Geometry of Normal Forms for Dynamical Systems}

\Author{Giuseppe Gaeta}

\Address{Dipartimento di Matematica, Universit\`a degli Studi di
Milano, via Saldini 50, 20133 Milano (Italy); e-mail: {\tt
giuseppe.gaeta@unimi.it}}

\begin{abstract}
  \noindent
  We discuss several aspects of the
geometry of vector fields in (Poincar\'e-Dulac) \emph{normal
form}. Our discussion relies substantially on \emph{Michel theory}
and aims at a constructive approach to simplify the analysis of
normal forms via a splitting based on the action of certain
groups. The case, common in Physics, of systems enjoying an
\emph{a priori} symmetry is also discussed in some detail. \hfill
\end{abstract}



\section{Introduction}

Most applications of Mathematics in Natural Sciences go through
Differential Equations. These are generically \emph{nonlinear},
and nonlinear differential equations as a rule cannot be solved.
Thus the only way to get some analytical information about their
behavior is through \emph{perturbation theory} \ii{perturbation
theory} -- in particular for systems which are in some sense close
to integrable (e.g. linear) ones, or for solutions which are in
some sense close to exactly known ones.

Henri Poincar\'e (1854-1912) set at the basis of Perturbation
Theory his method of \emph{normal forms} \ii{normal forms}
\cite{Poi1,Poi2} (for the life and work of Poincar\'e, see
\cite{VerPoi}). Here we will be concerned in particular with
normal forms for finite dimensional dynamical systems
\ii{dynamical systems}, but we stress that Poincar\'e's approach
also extends to \ii{evolution PDEs} evolution PDEs, see e.g.
\cite{ColEck,Eck,Niko}.

We will be specially interested in some \emph{geometric} feature
of the normal form approach; it is maybe worth stressing that the
motivation and goal for this has not to be traced to a desire of
mathematical abstraction, or to a preference for the geometric
(rather than analytic) approach, but rather to concrete
computational tasks. I hope I will convince the reader of the
advantages of having (also) a geometrical view of this topic.

The \emph{plan of the paper} is as follows. In Section
\ref{sec:NF} we give a short account of (the basics of) the normal
forms construction (which can be skipped by the reader having some
basic knowledge of this); in Section \ref{sec:symmNF} we discuss
the symmetry properties of systems in normal form. In Section
\ref{sec:Mic} we mention some feature of the Michel
theory\footnote{The name of Michel is associated to this theory in
particular for Physics applications; the mathematically oriented
reader will associate to it the names of Hilbert, Schwarz,
Procesi, Bredon, Bierstone, Thom among others. See also Sections
\ref{sec:Mic} and \ref{sec:discussion}, as well as
\cite{Fieldbook}.} \cite{Mic} of symmetric vector fields and
potential; these concerns how these object can be retraced to the
\emph{orbit space}, which -- when can be properly defined, see the
discussion there -- is in general a \emph{stratified manifold}
\cite{GorMcP}. We can then combine the two, which we do in Section
\ref{sec:MicNF}, discussing how the peculiar features of systems
in normal form allow, through the use of Michel theory and more
generally of invariants theory \cite{OlvInv} (and the separation
of vector fields in parts along and transversal to the group
action \cite{Chossat,ChoKoe,Koenig,Krupa}) to obtain a very
effective splitting.

In many relevant cases in Physics, the systems under study have
some symmetry property (e.g. under space rotations, Lorentz boost,
etc.); in this case the physical symmetry and the one built in the
normal form construction can combine in different ways, and the
procedure discussed so far can be further enhanced; this is
discussed in Section \ref{sec:symmetry} for general symmetries,
and in Section \ref{sec:linGNF} in the case where the physical
symmetries act \emph{linearly}. Under certain circumstances, the
symmetry properties enforce a \emph{finite} normal form, as
discussed in Section \ref{sec:finite}; and it may even happen that
-- again due to symmetry properties -- the normal form (or even
any symmetric vector field) has a \emph{gradient} structure, see
Section \ref{sec:grad}, which in turn may lead to
\emph{spontaneous linearization}, i.e. to dynamics being
asymptotically linear, see Section \ref{sec:linear}.

Needless to say, many topics remain outside this treatment; some
of these are briefly mentioned in the final Section
\ref{sec:discussion}, where we also summarize and discuss our
findings.

The paper is completed by several Appendices. In Appendix
\ref{sec:NFcomp} we recall the basic features of the normal forms
construction; Appendices \ref{sec:examples} and \ref{sec:Hopf} are
devoted to applications of the unfolding procedure described in
Section \ref{sec:MicNF}, respectively one to some illustrative
Examples and the other to the case of Hopf and Hamiltonian Hopf
bifurcations.

Finally, albeit our discussion will mostly be conducted at the
formal level, leaving the issue of converge of the involved series
(which in practical applications are of course always truncated to
some finite order) to a case-by.case discussion\footnote{The
theory is constructive, and all transformations are explicitly
determined; so one can determine explicitly also the radius of
convergence of the resulting (infinite or truncated) series of
transformations; see below.}, it makes of course a lot of sense to
try having information about this beforehand, i.e. in general
terms. This is the subject of Appendix \ref{sec:conv}.

\bigskip

I would like to stress two points concerning matters \emph{not}
discussed here:

\medskip\noindent (A) Another relevant case in
Physics is of course that where the system under study is
\emph{Hamiltonian}. The theory is of course well developed in this
case, and actually it has the advantage of dealing more
economically with a scalar object (the Hamiltonian) rather than
with a vector one (the dynamical vector field). We have chosen not
to deal specifically with this case, for two reasons: $(a)$
Hamiltonian vector fields are a special type of vector fields,
i.e. our general discussion will also cover the Hamiltonian case;
$(b)$ adding a specific discussion of the Hamiltonian case would
have made the article even longer, while this is already beyond
the limits assigned by the Editor, whom we thank both for the
invitation to contribute to this volume and for the patience in
this respect.

\medskip\noindent (B) Similar considerations apply to study of the
dynamics near a \ii{relative equilibrium} \emph{relative
equilibrium} (e.g. a periodic orbit or an invariant torus) rather
than a simple one; most of the approach and results described
below are extended to this more general setting, but this would
cause the article to grow far too much.
\bigskip

The symbol $\odot$ will mark the end of a Remark, while
$\triangle$ the end of a proof. Summation over repeated indices
will be routinely assumed, except in certain formulas where sums
are explicitly indicated.


\section{Normal forms}
\label{sec:NF}

Let us consider a dynamical system \beq \label{eq:DS} {\dot x} \ =
\ f (x) \eeq in a smooth $n$-dimensional manifold $M$. We assume
that there is some equilibrium point $x_0 \in M$, thus $f(x_0 ) =
0$, and as we are specially interested in the behavior near $x_0$,
we will consider a local chart with origin $x_0$ (so from now on
$x_0 = 0$), coordinates $x^i$ and Euclidean metric. We can thus
write \eqref{eq:DS} in components as \beq {\dot x}^i \ = \ f^i (x)
\ ; \eeq moreover, as we wish to study the situation nearby the
origin, we expand $f(x)$ in a Taylor series, and write \beq
\label{eq:DSx} {\dot x}^i \ = \ \sum_{k=0}^\infty f_k^i (x) \ ,
\eeq where the $f_k$ are homogeneous of degree $k+1$ in the $x$
(the reason for this ``notational shift'' will be apparent in the
following),
$$ f_k (a x) \ = \ a^{k+1} \ f_k (x) \ . $$
The term $f_0 (x)$ is linear, and will have a special role in the
following; we will also write \beq f_0 (x) \ = \ A \, x \ . \eeq
As well known, the constant matrix $A$ can always be decomposed
into a semisimple and nilpotent part (Jordan normal form), and the
two commute with each other:
$$ A \ = \ A_s \ + \ A_n \ ; \ \ [A_s , A_n] \ = \ 0 \ . $$

In the following, we will always assume (i) and (ii) below, and be
mostly interested in the case where (iii) also holds:

\medskip\noindent
{\tt (i)} $A_s \not= 0$;
\par\noindent
{\tt (ii)} The local coordinates have been chosen so that $A_s$ is
diagonal,
$$ A_s \ = \ \mathtt{diag} (\lambda_1 , ... , \lambda_n ) \ . $$
\par\noindent {\tt (iii)} $A_n = 0$.

\medskip\noindent {\bf Remark 1.} As well known, the eigenvalues of
$A$ coincide with those of $A_s$, hence (for the choice of
coordinates mentioned above) with the entries on the diagonal of
$A_s$. It is also well known that if all the $\lambda_i$ are
distinct, then necessarily $A_n = 0$; on the other hand, one can
have $A_n= 0$ even with multiple eigenvalues. For normal forms
with $A_n \not= 0$, see e.g. \cite{Gram,GramYosh}. \EOR

\medskip\noindent {\bf Remark 2.} Note that in many (but not all)
cases of physical interest, including the (special, but relevant)
case of Hamiltonian systems near a non-degenerate elliptic
equilibrium point, $A_n = 0$; moreover in that case the
eigenvalues $\lambda_i$ come in pairs of complex conjugate ones.
For the case of an elliptic equilibrium, we have purely imaginary
eigenvalues, $\lambda_m = \pm i \, \omega_k $. \EOR

\medskip\noindent {\bf Remark 3.} We stress that here we are
considering a \emph{given} $A$. For system depending on an
external parameter -- as e.g. those met in studying \emph{phase
transitions} -- it is more appropriate to consider families of
matrices $A(\mu )$ depending on such parameters. In cases of
interest for Physics, as indeed in phase transitions, these go
through $A=0$ (or at least $A_s = 0$) and hence our hypothesis, in
particular {\tt (i)} above, are necessarily violated. See in this
respect, and in a concrete physical application, the discussion in
\cite{GLandau}. \EOR
\bigskip

It is obvious that the linearized system \beq \label{eq:NFxlin}
{\dot x} \ = \ A \, x \eeq can be solved as $x(t) = \exp [ A t]
x_0$. One would expect that -- at least until $|x|$ remains small
-- the solutions to the full system \eqref{eq:DSx} are
approximated by those to \eqref{eq:NFxlin}. How good is this
approximation will of course depend on the nonlinear terms; in
particular we expect that if the first nonlinear terms
$f_1,f_2,...$ are actually vanishing, the approximation will be
better.

If we were able -- without altering the linear term -- to find
coordinates which would make the nonlinear terms vanishing up to
some finite but arbitrary order $N$, the solution $x(t) = \exp[A
t] x_0$ to the linear equation (in the new coordinates) would
approximate the full solution (in the new coordinates) with
arbitrary precision.

Poincar\'e showed that -- subject to a relevant \ii{non-resonance
condition} \emph{non-resonance condition} on the spectrum of
$A_s$, see below -- not only such changes exist\footnote{Here by
``exist'' we mean they exist \emph{formally}. More precisely, they
are described by a series, which is in general only formal;
criteria for the convergence of the series (at least in some small
neighborhood of the origin) have of course been widely studied,
see e.g. \cite{CWaam} and Appendix \ref{sec:conv} below.}, but can
be determined algorithmically \cite{Poi1,Poi2}.

The work by Poincar\'e was then extended by his pupil Henri Dulac
(1870-1955) who studied what happens when the non-resonance
condition is violated \cite{Dulac}; he showed that albeit in this case
nonlinear terms can in general not be eliminated, they can be
``simplified'' (in a sense to be explained below), or -- as one
now says -- \emph{normalized}. One can indeed reduce the system
(up to some finite but arbitrary order) to one which contains only
\emph{resonant} terms, in the sense to be discussed in a moment.

\medskip\noindent
{\bf Remark 4.} In the case of Hamiltonian vector fields one can
work directly on the Hamiltonian (one scalar function) rather than
on the associated vector field (with $n$ components, i.e. $n$
scalar functions). This situation was studied by George David
Birkhoff\footnote{Not to be mistaken with his son Garrett Birkhoff
(1911-1996); he is  also associated to Poincar\'e through the so
called (Capelli)-Poincar\'e-Birkhoff-Witt theorem.} (1884-1944)
for the non-resonant case \cite{Birk}, and by Fred Gustavson for
the resonant one \cite{Gustav}. \EOR
\bigskip

We will present the normal forms construction in an Appendix, for
the reader not already familiar with it (standard references for
it are \cite{ArnGMDE,Elp}. Here we will just characterize the
vector fields which are obtained as a result of the normalization
procedure.



\medskip\noindent
{\bf Definition 1.} {\it A vector with components $x^\mu {\bf
e}_i$ is of order $m$ if $\mu_1 + ... + \mu_n = m$, and it is
\emph{resonant} (with $A_s$) if \beq \label{eq:resonant} \mu \cdot
\lambda \ = \ \sum \mu_k \, \lambda_k \ = \ \lambda_i \ . \eeq A
vector field $X_f = f^i \pa_i$ is resonant  (with $A_s$) if its
components are resonant.}
\bigskip

Obviously the set of resonant vectors of a given order is a linear
space of finite (possibly zero) dimension, and the same holds --
except for the finite dimension, in general -- if we consider
resonant vectors of any order. Our discussion
in Appendix \ref{sec:NFcomp} can be summarized as

\medskip\noindent
{\bf Proposition 2.} {\it Let the vector field $X_f = f^i \pa_i$
admitting a zero in the origin have linear part $X_0 = (A x)^i
\pa_i$, with $A = A_s$. Then $X_f$ is in normal form if and only
if it is resonant with $A_s$.}

\section{Normal forms and symmetry}
\label{sec:symmNF}

The discussion of the previous section allows to characterize
(vector fields in) normal forms in terms of their \emph{symmetry
properties}: in fact, the vector fields $X_k$ associated to all
the nonlinear terms $F_k$ do commute with $X_0$, the one
associated to the linear part\footnote{We recall once again we are
assuming, for the sake of simplicity, that this linear part is
semisimple, $A=A_s$.} of the system: \beq \label{eq:res} X_k \ :=
\ F_k^\a (x) \ \frac{\pa}{\pa x^\a} \ ; \ \ \left[ X_0 , X_k
\right] \ = \ 0 \ . \eeq This condition also provides a
characterization of \ii{resonant vector fields} \emph{resonant
vector fields}.\footnote{At first sight this is an invariant,
i.e.coordinate-independent, characterization. However, note that
it depends on what is the linear part of the dynamics, and this is
dependent on the choice of coordinates, albeit will not change
under well-behaved coordinate changes.}

It is immediate to observe that \eqref{eq:res}, together with the
Jacobi identity, implies that:

\medskip\noindent
{\bf Lemma 1.} {\it Vector fields which are resonant with a given
linear one $X_0$, span a Lie algebra.}
\bigskip

It may be useful to consider the transposition of the Lie bracket
(i.e. the commutator) in terms of components of the vector fields.
This is the Lie-Poisson bracket between vector functions $f,g :
{\bf R}^n \to {\bf R}^n$, defined as \beq \label{eq:LP} \{ f , g
\}^i \ := \ (f^j \pa_j) , g^i \ - \ (g^j \pa_j ) \, f^i \ ; \eeq
equivalently, \beq \label{eq:LP2} \{ f , g \} \ := \ (f \cdot
\nabla) \, g \ - \ (g \cdot \nabla) \, f \ . \eeq It is immediate
to check that if $ X = f^i \pa_i$, $Y = g^i \pa_i$, then \beq Z \
= \ [X,Y] \ = \ h^i \pa_i \ ; \ \ \ h \ = \ \{ f , g \} \ . \eeq

We will thus consider the set $V \equiv V^{(A)}$ of (polynomial)
equivariant vector functions, i.e. of functions $f : {\bf R}^n \to
{\bf R}^n$ such that \beq \{ Ax , f \} \ = \ 0 \ . \eeq In
particular, we will consider the (linear space) of equivariant
functions homogeneous of degree $k+1$, denoted as $V_k$.

It will also be natural to consider the ring $I \equiv I^{(A)}$ of
(polynomial) scalar functions invariant under the linear part of
$X$, i.e. functions $\b : {\bf R}^n \to {\bf R}$ such that $X_0
(\b) = 0$. In particular, we will consider the (linear space) of
invariant functions homogeneous of degree $k$, denoted as $I_k$.

It is rather obvious that $V^{(A)}$ has the structure of a \ii{Lie
module} \emph{Lie module} over $I^{(A)}$. It is also obvious that
for any $\b \in I_m $, $f \in V_k$, we have $\b f \in
\mathcal{V}_{k+m}$. Further details are provided e.g. in Chapter
III of \cite{CGbook}. A full characterization of normal forms in
terms of symmetry is provided by the following Lemma; see
\cite{WalNF} for its proof:

\medskip\noindent
{\bf Lemma 2.} {\it If $A = (Df)(0)$, then the NF $\widehat{f}$
for $f$ can be written in the form \beq \widehat{f} (x) \ = \
\sum_{j=0}^s \mu_j (x) \ M_j \, x \ , \eeq where $M_j$ are a basis
for the linear space of real $n$-dimensional matrices commuting
with $A$, and $\mu_j (x)$ are scalar (polynomial or possibly
rational) functions for the linear flow $\dot{x} = A x$.}

\medskip\noindent
{\bf Remark 5.} With the notation used in this Lemma, it is
natural to choose one of the $M_j$, say $M_0$, to coincide with
$A$; note that $s \ge 0$, and that if $s=0$ we have $\widehat{f}
(x) = [1 + \a (x)] A x $ with $\a$ an invariant function; we are
thus in the framework of what is known as ``condition $\a$'', see
Appendix \ref{sec:conv}. \EOR
\bigskip

Summarizing, vector fields in normal form are characterized by
their symmetry property under the vector field $X_0$. Recalling
that normal forms are by definition also polynomial, the task of
describing the most general normal form -- and its dynamics -- for
a given linear part is then reduced to the task of studying the
most general polynomial vector field commuting with (hence
covariant w.r.t.) a given linear one.

\section{Michel theory}
\label{sec:Mic}

We are thus led to consider, in full generality, polynomial (or,
for that matter, $C^\infty$) vector fields which commute with a
given linear one $X_0$; equivalently, which are symmetric (that
is, equivariant) under the action of a linear Lie group $G_0$,
generated by $X_0$.

It should be noted that in many physical situations the system
(even before the reduction to normal form) will be required to
have some symmetry properties on the basis of the Physics it
describes. The more common ones are of course symmetries under
translations, rotations and inversions (Euclidean group) or in the
relativistic context under the Lorentz or the full Poincar\'e
group.

Moreover, it may happen that the original system \eqref{eq:DS} has
some special symmetry beyond (or instead of) those mentioned
above; in full generality we assume that the original system has a
Lie symmetry described by a group $G$ with Lie algebra
$\mathcal{G}$. In this case it is well known that the whole
normalization procedure an be performed preserving such
symmetries, see e.g. \cite{CGbook} and references given there. The
normal form will then correspond to vector fields which are
symmetric under $G$ \emph{and} also under $G_0$\footnote{Note that
it may happen that $G_0 \subset G$. E.g., consider the case where
we have a dynamical system in ${\bf R}^2$ required to be
rotationally invariant and whose linear part is just a rotation.}

Thus we consider vector fields in ${\bf R}^n$  which are
$G$-equivariant for $G$ a general Lie group acting in ${\bf R}^n$.


We will consider the \ii{orbit space} \emph{orbit space} $\Omega =
M/G$. Its elements are the $G$-orbits $\omega$ in $M$, i.e. the
sets \beq \omega_x \ = \ \{ y \in M \ : \ y = g x \ \mathrm{for \
some \ } g \in G \} \ = \ G x \ . \eeq Note that here (and below)
we think the $G$-action (i.e. the representation $T$ through which
$G$ acts) in $M$ to be given, and identify $g x$ with $T_g x$,
etc.

The distance between two orbits $\omega_x$ and $\omega_y$ is
defined as
$$ \delta (\omega_x , \omega_y ) \ = \ \min \left[ d (\xi, \eta) , \ \xi \in
\omega_x , \ \eta \in \omega_y \right] $$ with $d$ the standard
distance in $M$.

\medskip\noindent
{\bf Assumption.}
We will from now on assume that $G$ acts \emph{regularly} in $M$.

\medskip\noindent
{\bf Remark 6.} This is automatically satisfied if $G$ is a
\emph{compact} Lie group, but typically $G_0$ (see notation above)
is not compact, at least for generic dynamical systems (we have a
compact $G_0$, actually $G_0 = {\bf T}^\ell$, for the special but
relevant case of a Hamiltonian system near an elliptic fixed
point). \EOR

\medskip\noindent
{\bf Remark 7.} In many respects, the requirement of a compact
group $G$ can be replaced by a weaker one, i.e. that $D(H):=
N(H)/H$ is compact for maximal isotropy subgroups $H \subseteq G$.
Here we denote by $N(H) = N(H,G)$ the normalizer of $H$ in $G$,
that is the greater subgroup of $G$ in which $H$ is a normal
subgroup; obviously $H \subseteq N(H)$. Then $D(H) := N(H)/H$, the
quotient being well defined since $H$ is actually by definition a
normal subgroup in $N(H)$. \EOR
\bigskip

With this hypothesis, $\delta(\omega_x , \omega_y ) = 0$ if and
only if $x,y$ belong to the same orbit, i.e. if and only if $y = g
x$ for some $g \in G$. (A counterexample when the assumption is
not satisfied is provided, as usual, by the irrational flow on the
torus.)

The orbit space $\Omega = M/G$ is then a \ii{stratified manifold}
\emph{stratified manifold} \cite{GorMcP} in the sense of Algebraic
Geometry, i.e. the union of smooth manifolds with manifolds of
smaller dimension lying at the border of those of greater
dimension.\footnote{Note that this does \emph{not} imply that
$\Omega$ itself is a manifold. A familiar example of a stratified
manifold which is not a manifold is provided by a cube. The
interior of the cube is a three-dimensional manifold $M^3$, the
(interior of the) faces are two-dimensional manifolds $M^2 \subset
\pa M^3$, the (interior of the) edges are one-dimensional
manifolds $M^1 \subset \pa M^2$, and the vertices are
zero-dimensional manifolds $M^0 \in \pa M^1$.}

There is also a different (in principles) stratification of
$\Omega$, based on symmetry properties; we will refer to this as
its \ii{isotropy stratification} \emph{isotropy stratification}.

Given the $G$-action on $M$, we can associate to any $x \in M$ its
isotropy subgroup \beq G_x \ = \ \{ g \in G \ : \ g x = x \} \
\subseteq \ G \ . \eeq It is quite clear that points on the same
$G$-orbit have isotropy subgroups which are conjugated in $G$. In
fact, if $y = g x$ with $g \in G$, then $G_y \ = \ g \, G_x \,
g^{-1}$.

If we define an \ii{isotropy type} \emph{isotropy type} as the set
of points in $M$ which have isotropy subgroups which are
$G$-conjugated, it is then clear that points on $\omega_x$ all
belongs to the same isotropy type $[G_x]$, and we can assign to
$\omega = \omega_x$ an isotropy class (i.e. an isotropy type).

There is a natural inclusion relation among (isotropy) subgroups
of $G$, and this also naturally extends to a relation among
isotropy types: we say that $[G_y] \subseteq [G_x]$ if there are
subgroups $G_1 \in [G_y]$ and $G_2 \in  [G_x]$ such that $G_1
\subseteq G_2$.

Then we can stratify $\Omega$ (and actually also $M$) on the basis
of the isotropy properties of orbits $\omega \in \Omega$; there
will be a generic stratum with lower isotropy $G_0$ (usually --
and surely if the $G$-action in $M$ is effective -- just $G_0 = \{
e \}$), and then higher and higher strata with isotropy types
corresponding to larger and larger isotropy subgroups of
$G$.\footnote{We stress that while the subgroups of $G$ are
defined independently of the way $G$ acts in $M$, the lattice of
isotropy subgroups depends on the $G$-action. For example, if $G$
acts via the trivial representation, all points have isotropy
$G$.}

It was realized by L.Michel \cite{Mic0,Mic,MicPR} that the
geometric stratification of $\Omega$ (this is also called its
\emph{Whitney stratification}) is coherent with its \emph{isotropy
stratification}.\footnote{E.g., if we consider $R^3$ and on it $G
= Z_2 \times Z_2 \times Z_2$ acting as $R_x \times R_y \times
R_z$, where $R_\a$ is the reflection in the reflection in the
coordinate $\a$, the orbit space $\Omega$ is made of a octant in
$R^3$, say the first one. Points with three non-zero coordinates
have isotropy type $[e]$, points on one of the faces, say the one
with the $\a$ coordinate equal to zero,  have isotropy type $[R_\a
]$; points on one of the edges, say the one with both $\a$ and
$\b$ coordinate equal to zero, have isotropy type $[R_\a \times
R_\b ]$; and the vertex in the origin has full isotropy $[G]$.}

In the case  of an equivariant dynamics, i.e. where a
$G$-covariant vector field $X$ is defined in $M$, this has a very
relevant consequence. That is, the vector field is everywhere
tangent to strata in $M$, and hence: (i) it can be projected to a
vector field $X_\omega$ in $\Omega$; (ii) strata in $M$ and in
$\Omega$ are invariant under the flow defined by $X$ and
respectively $X_\omega$.\footnote{A finer analysis in this respect
is contained in \cite{ChoKoe,Koenig,Krupa}; see also
\cite{Chossat} for a comprehensive discussion.}

Thus, we conclude that symmetric dynamics can -- under rather mild
conditions on the geometry (topology) of the relevant group action
-- be projected to the orbit space. Needless to say, this is in
general of smaller (sometimes much smaller) dimension and hence
hopefully more easily studied. We will see in the following that
actually \emph{if} we are able to solve this ``simpler'' -- but
nevertheless in general \emph{nonlinear} -- dynamics, the dynamics
of systems in normal form can be reconstructed by solving
\emph{linear} (albeit non-autonomous) equations.

\medskip\noindent
{\bf Remark 8.}  A very readable introduction to Michel theory and
its (original) physical applications is provided by Abud and
Sartori \cite{AbuSar} (see also \cite{Sartori}); for a more
comprehensive discussion, see \cite{MicPR}. For an extension to
\ii{gauge theories} \emph{gauge theories}, see \cite{GMAP} (the
geometry of gauge orbit space is discussed e.g. in
\cite{ACM,Kond}). Dynamical aspects are discussed in
\cite{Field96,FR,Fieldbook} and in
\cite{Chossat,ChoKoe,Koenig,Krupa}. For the original issues
leading physicists to consider these problems, see
\cite{CabMai,Mic0,MicRad}. The work of R. Palais on the
\ii{symmetric criticality principle} ``symmetric criticality
principle'' \cite{Pal1,Pal2,Pal3} could be seen as an attempt to
extend this theory to the infinite dimensional case. \EOR

\medskip\noindent
{\bf Remark 9.} As for the mathematical aspects of (or counterpart
to) Michel theory, this would lead to a long discussion, and we
will just refer to \cite{Field,Fieldbook}. \EOR

\section{Unfolding of normal forms}
\label{sec:MicNF}

We can now go back to dynamical systems in (Poincar\'e-Dulac)
normal form. The idea we want to pursue is to \emph{increase} the
dimension of the system by embedding it into a larger system
carrying the same information. The goal is to have a larger system
with simpler properties (this idea was successfully carried on in
the famous paper by Kazhdan, Konstant and Sternberg \cite{KKS} on
the Calogero integrable system \cite{Cal}); in this context we
refer to the larger system as an \ii{unfolding} \emph{unfolding}
of the original one.

In order to do this, we should look more carefully into
resonances. These can be of two types, i.e. those corresponding to
\ii{invariance relations} \emph{invariance relations} on the one
hand, and \ii{sporadic resonances} \emph{sporadic resonances} on
the other. We now define these concepts.

\subsection{Resonances, invariance relations, sporadic resonances}

Recalling that we denote by $\lambda_i$ the eigenvalues of $A_s$
(see section \ref{sec:NF}), it is clear that if there are
non-negative integers $\sigma_i$ such that \beq \label{eq:invrel}
\sum_{i=1}^n \ \sigma_i \ \lambda_i \ = \ 0 \ , \eeq say with
$s_1+... \sigma_n = |\sigma | \not= 0$, these $\sigma_i$ can
always be added (term by term) to any resonance vector $\mu_i$ (of
order $|\mu |$) to produce new resonant vectors (of order $|\mu |
+ k |\sigma |$, with any $k \in {\bf N}$).

We say that \eqref{eq:invrel} identifies an \emph{invariance
relation}. Having invariance relations is the only way to have
infinitely many resonances (and hence infinitely many terms in a
normal form) in a finite dimensional system \cite{WalNF}.

Any nontrivial resonance \eqref{eq:resonant} such that there is no
$\sigma$ with $\sigma_i \le \mu_i$ (for all $i = 1,...,n$)
providing an invariance relation, is said to be a \emph{sporadic
resonance}. Sporadic resonances are always in finite number
(possibly zero) in a finite dimensional system \cite{WalNF}.

\subsection{Invariance relations and invariant functions}

It is clear that invariance relations are associated to invariant
functions under the action of $G_0$, or equivalently of (its
generator, i.e.) the linear vector field $X_0 = (A_s x) \nabla$;
and conversely, any polynomial scalar function which is invariant
under $G_0$ is associated to an invariance relation. In fact, if
$$ I(x) = x_1^{\sigma_1} ... x_n^{\sigma_n} \ , $$ we immediately have
$$ X_0 (I) \ = \ \lambda_i x_i \ \frac{\pa}{\pa x_i} I(x) \ = \
\left( \sum_{i = 1}^n \sigma_i \, \lambda_i \right) \ I (x) \ = \
0 \ .
$$ We assume there are $r \ge 0$ independent invariance relations
(here ``independent'' means that $I_1 (x) ,... , I_r (x)$ are
functionally independent).

If we look at the full dynamics of $I(x)$, it follows from
$[X,X_0] = 0$ that $I(x)$ remains always $G_0$-invariant; hence
when we write \beq \frac{d I (x)}{d t} \ = \ \frac{\pa I (x)}{\pa
x_i} \ \frac{d x_i}{d t} \ = \ \frac{\pa I (x)}{\pa x_i} \ f^i (x)
\ := \ Z (x) \eeq the function $Z(x)$ must be itself invariant
under $G_0$; for what we have said above, this means $Z(x)$ can be
written as \beq Z(x)\ = \ \Phi [I_1 (x) , ... , I_r (x) ] \ , \eeq
i.e. that the set of generators for the ring of $G_0$-invariant
functions evolves in time according to \beq \label{eq:invevo}
\frac{d I_a (x)}{d t} \ = \  \Phi_a [I_1 (x), ... , I_r (x)] \ \ \
\ (a = 1,...,r) \ . \eeq

Thus if we introduce auxiliary variables $\varphi_a$ ($\a =
1,...,r$) and let them evolve according to \beq
\label{eq:invevophi} \frac{d \varphi_a }{d t} \ = \  \Phi_a
(\varphi_1 , ... , \varphi_r ) \ , \eeq the relation $\varphi_a =
I_a (x_1 , ... , x_m )$ -- if satisfied at $t=0$ -- will be
preserved by the flow.

\subsection{Sporadic resonances and auxiliary variables}

We will now consider auxiliary variables $w_i$ associated to
sporadic resonances \cite{GGlmp}; if \eqref{eq:resonant}
identifies a sporadic resonance, we define $$ R_i \ = \
x_1^{\mu_1} ... x_n^{\mu_n} \ . $$ In this case we immediately
have
$$ X_0 (R_i) \ = \ \left( \sum_k \lambda_k \mu_k \right) \ R_i \ = \ \lambda_i \, R_i \ . $$
Thus the functions $R_i [x (t)]$ are covariant, in the sense they
evolve as the $x_i$ involved in the (sporadic) resonance relation.
Note in particular this means they evolve \emph{linearly} --
despite being \emph{nonlinear} functions of the $x$. Moreover,
again assuming $R$ is associated to \eqref{eq:resonant} and
considering the linear dynamic, by construction
$$ X_0 [ x_i - R_i(x) ] \ = \ \lambda_i \, x_i \ - \
\sum_k \frac{\pa R_i}{\pa x_k} \, \lambda_k x_k \ = \ \lambda_i \,
x_i \ - \ \left( \sum_k \lambda_k \mu_k \right) R_i \ = \
\lambda_i \ (x_i \ - \ R_i ) \ . $$ In particular, the manifold
identified by
$$ x_i \ = \ R_i (x) $$ is by construction invariant.

\medskip\noindent
{\bf Remark 10.} Note that we can have two different sporadic
resonances (with $\mu$ and $\mu' \not= \mu$) involve the same
distinguished variable $x_i$ only if they are actually related by
an invariance relation, as follows immediately from noting that
$\mu' - \mu = 0$. Thus we can have at most one independent
sporadic resonance for degree of freedom, and the notation $w_i$
is convenient to identify the distinguished variable involved in
this. \EOR
\bigskip

We will let these auxiliary variables $w_i$ -- which we see as
independent variables -- evolve according to \beq \label{eq:wevo}
\frac{d w_i}{d t} \ = \ \frac{\pa R_i}{\pa x_k} \ \frac{d x^k}{d
t} \ = \ \frac{\pa R_i}{\pa x_k} \ f^k (x) \ . \eeq

\subsection{Unfolding of normal form}

Summarizing, we have three types of variables: the $n$ natural
coordinates $x^i$, the $m \le n$ auxiliary variables $w^i$
associated to sporadic resonances, and the $r$ auxiliary variables
$\phi^a$ associated to invariance relations. Correspondingly, we
have a dynamics in a $(n+m+r)$-dimensional space, whose general
form is
\begin{eqnarray}
\dot{x}^i &=& f^i (x,w,\phi) \nonumber \\
\dot{w}^i &=& g^i (x,w,\phi) \label{eq:genufform} \\
\dot{\phi}^a &=& h^a (x,w,\phi) \ . \nonumber \end{eqnarray}

In order for this to represent our original dynamics
\eqref{eq:DS}, the functions $f,g,h$ should be suitably assigned.
Obviously $f$ should reproduce the $F$ appearing in \eqref{eq:DS},
and $g$, $h$ should be compatible with the identification of $w$,
$\phi$ given by \eqref{eq:invevo}  and \eqref{eq:wevo}
respectively. Moreover, precisely these identifications introduce
some ambiguity in the writing of $f,g,h$ in terms of the enlarged
set of variables. This ambiguity can be used to write the
equations in a convenient form -- which is precisely the reason to
introduce the auxiliary variables.

In fact, we have the following result, which is a restatement of
those given in our previous work \cite{GWjnmp}.

\medskip\noindent
{\bf Lemma 3.} {\it The function $h$ can be written as an analytic
function of the $\phi$ variables alone; the functions $f$ and $g$
can be written as analytic functions, linear in the $x$ and the
$w$ variables.}

\medskip\noindent
{\bf Proof.} Let us start by considering $f$; the function $F^s
(x)$ appearing in \eqref{eq:DS} is made of resonant terms only,
and these can always be written in terms of invariance relations
and sporadic resonances as
$$ F^s (x) \ = \ \a_s [\phi_1 (x), ... \phi_r (x)] \, x_s \ + \
\b_s [\phi_1 (x), ... \phi_r (x)] \,
w_s\ ; $$ thus it suffices to define (no sum on $i$) \beq f^i
(x,w,\phi) \ = \ \a_i (\phi) x_i \ + \ \b_i (\phi ) w_i \ . \eeq

Let us now consider the $g^i$. In this case we would have
$$ \dot{w}^i \ = \ \frac{\pa w^i}{\pa x^j} \  \dot{x}^j \ = \
\frac{\pa w^i}{\pa x^j} \ f^j (x,w,\phi) \ ; $$ but all terms on
the r.h.s. are resonant with $x^j$, hence can be written in terms
of resonant monomials and $x^j$, or $w^j$ itself.

Finally, we have seen above that the time evolution of $\phi^a$ is
written in terms of invariant functions only, hence of the $\phi$
themselves. \hfill $\triangle$
\bigskip

This is a remarkable result in that it allows to identify the main
obstacle to the analysis of systems in normal form and tells how
to proceed in this task. We will write its relevant consequences
in the form of a Corollary.

\medskip\noindent
{\bf Corollary.} {\it The equations \eqref{eq:genufform} can be
written as
\begin{eqnarray}
\dot{x}^i &=& F^i_{\ j} (\phi) \ x^j \nonumber \\
\dot{w}^i &=& G^i_{\ j} (\phi) \ w^j \label{eq:redufform} \\
\dot{\phi}^a &=& h^a (\phi) \ . \nonumber \end{eqnarray} If the
(in general, nonlinear) last set of equations is solved, providing
$\phi = \phi (t)$, then the first two sets reduce to
\begin{eqnarray}
\dot{x}^i &=& \widehat{F}^i_{\ j} (t) \ x^j \nonumber \\
\dot{w}^i &=& \widehat{G}^i_{\ j} (t) \ w^j
\label{eq:redredufform}
\end{eqnarray}
where of course $\widehat{F} (t) = F [\phi (t)]$, $\widehat{G} (t)
= G[\phi (t)]$, and we only have to solve \emph{linear} (in
general, non autonomous) equations.}
\bigskip

Some Examples of application of our construction are given in
Appendix \ref{sec:examples}. Application to the analysis of the
Hopf and the Hamiltonian Hopf bifurcations is given in Appendix
\ref{sec:Hopf}.

\medskip\noindent
{\bf Remark 11} Further developments of this approach are
discussed in \cite{GGSW,GWjde,GroWal}, see also \cite{SW15}; the
reader is referred to the original papers for detail. \EOR
\bigskip

%
%

\section{Normal forms in the presence of symmetry}
\label{sec:symmetry}

In the case where the original system \eqref{eq:DS} has some
symmetry -- possibly, but not necessarily, dictated by the Physics
it describes (e.g. covariance under rotation or the Lorentz group)
-- it is well known that the whole Poincar\'e-Dulac normalization
procedure can be carried out remaining within the class of
covariant objects: the Poincar\'e transformations at each step
will have covariant generating functions, and the normalized
vector fields will be covariant. Moreover, the vector fields in
normal form will have the extra symmetry defined by
$X_A$.\footnote{We stress this can induce a larger reduction, see
Example 2 and Remark \ref{sec:examples}.1 in Appendix
\ref{sec:examples}.}

In this case, we do \emph{not} have to study the most general
resonant vector field, but the most general \emph{resonant and
covariant} one. Needless to say, this is in general much less
general than requiring just resonance, i.e. in general a covariant
normal form will be simpler (in the sense of admitting less terms)
than a generic one.

The discussion of Section \ref{sec:MicNF} and the construction
described there still apply, except that now the role of $G_0$ is
played in general by a larger group $G$ (in practice, this is most
often a group with a \emph{linear} action; but this is not
necessarily the case). In particular, the invariant functions
$\phi^a$ will now be the functions which are invariant under
\emph{both} the linear dynamical group $G_0$ \emph{and} the group
$G$ of ``physical'' symmetries (in other words, only $G$-symmetric
invariance relations will have a role). Similarly, only sporadic
resonances which respect the $G$ symmetry will correspond to
resonant terms present in the normal form and hence to relevant
auxiliary variables $w^i$.

\medskip\noindent
{\bf Remark 12.} From this point of view, we should stress that
$G_0$ acts in general as a group of \emph{non-linear}
transformations, but its generator is associated to a linear
function of the (adapted) variables. That is, if $A =
\mathrm{diag}(\lambda_1 , ... \lambda_n )$ and $s$ is the group
parameter, then
$$ g \ = \ \exp [s A] \ : \ (x_1,..,x_n) \ \to \ \left( e^{s \lambda_1} x_1 ,
... , e^{s \lambda_n} x_n \right) \ . $$ Having a generator which
depends linearly on the $x$ simplifies in many ways the situation
to be studied. \EOR
\bigskip

Thus the extension to the symmetric case is essentially trivial
from the theoretical point of view; but it can lead to relevant
simplifications in practice. This is possibly better illustrated
by considering directly a concrete example, related to one of
those considered above.

\section{Normal forms and classical Lie groups}
\label{sec:linGNF}

In many physical applications, one meets systems with a symmetry
described by simple compact Lie groups, and in particular by the
\ii{classical groups} classical groups.

The relevant point here is that if the ``physical'' symmetry $G$
acts regularly (which is definitely the case for a linear
representation of a compact Lie group), we can forget about the
non-compact nature of the $G_0$ action, and perform reduction to
orbit space \emph{only} under the $G$-action.

This amounts to consider general (polynomial) vector fields which
are covariant under the $G$-action, and these can be studied in
general terms for the \ii{simple Lie groups} simple Lie groups.

The basic classification result here is the (general version of)
\ii{Schur Lemma} Schur Lemma and a simple consequence of this,
which we quote from Kirillov \cite{Kir}.

\medskip\noindent
{\bf Lemma 4 (Schur Lemma).} {\it Let the dimension of the
irreducible group representation $T$in a linear space over the
field ${\bf K}$ be at most countable; denote by $C(T)$ the
centralizer of $T$, by $c(T)$ the intertwining number $c(T) =
\mathrm{dim}_{\bf K} [C (T)]$. Then if ${\bf K} = {\bf C}$, $C(T)
\simeq {\bf C}$, $c(T) = 1$; if ${\bf K} = {\bf R}$, $C(T)$ is
isomorphc to either ${\bf R}$ or ${\bf C}$ or ${\bf H}$ and
correspondingly $c(T) = 1,2,4$.}
\bigskip

In the case ${\bf K} ={\bf R}$, the representation $T$ is said to
be of real, complex or quaternionic type according to the form of
$C(T)$, see above.

\medskip\noindent
{\bf Lemma 5.} {\it Let $T_{{\bf C}}$ be the complexification of
the real irreducible representation $T$. If $T$ is of real type
then $T_{{\bf C}}$is irreducible; if $T$ is of complex type then
$T_{{\bf C}}$ is the sum of two inequivalent irreducible
representations; if $T$ is of quaternionic type then $T_{{\bf C}}$
is the sum of two equivalent irreducible representations.}
\bigskip

We can thus classify symmetric normal forms in $R^n$ according to
the type of the irreducible representation describing the symmetry
\cite{Gsimp}. Note that if we have an irreducible
\emph{orthogonal} representation, this is necessarily transitive
on the unit sphere $S^{n-1}$ of the carrier space $R^n$, hence the
only invariant is $r$; as we want a \emph{polynomial} invariant we
should consider $\rho = r^2 = x_1^2 +... + x_n^2$. In this case
polynomial covariant vector fields are of the form \beq
\label{eq:tepvf} \dot{x} \ = \ \sum_{j=0}^s \ \mu_j (\rho) \ K_j \
x \ , \eeq where the $K_j$ are a basis for the set $C(T)$ of
$n$-dimensional real matrices commuting with $T$; we can and will
always choose $K_0 = I$.

\subsection{Real type}

In this case $C(T) \simeq R$, hence is given by multiples of the
identity, and \eqref{eq:tepvf} is just \beq \dot{x} \ = \ \mu_0
(\rho) \ x \ ; \eeq obviously this evolves towards spheres with
radius $\rho^*$ corresponding to the zeros of $\mu_0 (\rho)$; more
precisely towards those with $\mu_0 (\rho^*) = 0$, $\mu_0'
(\rho^*) < 0$.

In more detail, $\rho$ evolves according to
$$ \dot{\rho} \ = \ 2 \ \rho \ \mu_0 (\rho ) \ ; $$
note this is a separable equation and can hence be solved
computing a rational (as $\mu_0$ is a polynomial) integral,
$$ \int \frac{1}{\rho \ \mu_0 (\rho ) } \ d \rho \ = \ 2 \ (t - t_0) \ .  $$

The linear part is a multiple of the identity: no resonances are
present, and the normal form is just linear. Note that the
\ii{Poincar\'e criterion} Poincar\'e criterion applies (see
Appendix \ref{sec:conv}), thus there is a convergent normalizing
transformation.

\subsection{Complex type}

In this case $C(T) \simeq {\bf C}$; in other words we have two
independent matrices commuting with $T$, one of them is of course
the identity $I$, while the other will be denoted as $J$. Note
that $T$ is irreducible over $R$ but as a complex representation
it will be given by $T = T_0 \oplus \widehat{T}_0$, by \ii{Schur
lemma} Schur lemma. This implies $n = 2 m$.

Now, using coordinates adapted to this decomposition,
\eqref{eq:tepvf} reads just \beq \dot{x} \ = \ \mu_0 (\rho) \ I \,
x \ + \ \mu_1 (\rho ) \ J \, x \ ; \eeq in these coordinates, $J$
is written in block form as (the standard symplectic matrix)
$$ J \ = \ \pmatrix{0 & - I \cr I & 0 \cr} \ . $$

Now the linear part reads
$$ A \ = \ c_0 \ I \ + \ c_1 \ J \ ; $$
different cases are possible depending on the vanishing of the
constants $c_0$ and $c_1$. We exclude the fully degenerate case
$c_0 = 0 = c_1$, where we have $A=0$.

\begin{enumerate}

\item If $c_1= 0$, $c_0 \not= 0$, we are in the same situation as
in the real case: the linear part is a multiple of the identity
and no resonance is present; the normal form is linear (with a
convergent normalizing transformation).

\item If $c_0 \not= 0$, $c_1 \not= 0$, the eigenvalues of $A$ are
equal to $\lambda_\pm = c_0 \pm i c_1$ (each of these with
multiplicity $m$). Again no resonances are present, hence the
normal form is linear, and again the spectrum belongs to a
Poincar\'e domain and hence there is a convergent normalizing
transformation.

\item If $c_0 = 0$, $c_1 \not= 0$, the eigenvalues of $A$ are $
\lambda_\pm = \pm i c_1$ (each with multiplicity $m$). In this
case there is an invariance relation $\lambda_+ + la_- = 0$, hence
an infinite number of resonances. Moreover the spectrum does
\emph{not} belong to a \ii{Poincar\'e domain} Poincar\'e domain,
hence we are not guaranteed there exist a convergent normalizing
transformation.

\end{enumerate}

\subsection{Quaternionic type}

The only fundamental representation of a simple Lie group
realizing this case occurs for $G =SU(2)$, i.e. the \ii{quaternion
group} quaternion group itself; we thus discuss directly this case
in concrete terms.\footnote{There are higher representation of
other simple Lie groups of this type, see e.g. \cite{CG88}, but
these appear to be of little physical interest.}

The basis matrices of the Lie algebra $su(2)$ can be taken to be
$$  H_1 = \pmatrix{0&0&1&0\cr 0&0&0&1\cr -1&0&0&0\cr  0&-1&0&0\cr} \ , \
H_2 = \pmatrix{0&0&0&-1\cr 0&0&1&0\cr 0&-1&0&0\cr 1&0&0&0\cr} \ ,
\ H_3 = \pmatrix{0&-1&0&0\cr 1&0&0&0\cr 0&0&0&1\cr 0&0&-1&0\cr}
\ . $$ With these, $C(T)$ is spanned by the identity $I = K_0$ and
by the three matrices
$$ K_1 = \pmatrix{0&1&0&0\cr -1&0&0&0\cr 0&0&0&1\cr 0&0&-1&0\cr} \ , \
K_2 = \pmatrix{0&0&0&1\cr 0&0&1&0\cr 0&-1&0&0\cr -1&0&0&0\cr}  \ ,
\ K_3 = \pmatrix{0&0&1&0\cr 0&0&0&-1\cr -1&0&0&0\cr 0&1&0&0\cr} \
; $$ these do of course span another, not equivalent, $su(2)$
representation (see again the \ii{Schur Lemma} Schur Lemma).

The general form of \eqref{eq:tepvf} is thus \beq \dot{x} \ = \
\sum_{j=0}^3 \mu_j (\rho) \ K_j \, x \ . \eeq The linear part of
this is $A = \sum c_j K_j$, and we write
$$ \omega \ = \ \sqrt{c_1^2 + c_2^2 + c_3^2} \ . $$

The eigenvalues of $A$ are $\lambda_\pm = c_0 \pm  i \omega$, each
with multiplicity two. Several subcases are possible, as in the
complex case (we again exclude the fully degenerate case $c_0 = 0
= \omega$, where we have $A=0$).

\begin{enumerate}

\item If $\omega= 0$, $c_0 \not= 0$, the linear part is a multiple
of the identity; the normal form is linear with a convergent
normalizing transformation.

\item If $c_0 \not= 0$, $\omega \not= 0$, no resonances are
present, the normal form is linear, and there is a convergent
normalizing transformation.

\item If $c_0 = 0$, $\omega \not= 0$, then $ \lambda_\pm = \pm i
\omega$. There is an invariance relation, hence an infinite number
of resonances, and we are not guaranteed there exist a convergent
normalizing transformation.

\end{enumerate}

\medskip\noindent
{\bf Remark 13.} In all the three ($R,C,H$) cases, one can discuss
in rather general terms \ii{further normalization} \emph{further
normalization} \cite{GaePRF} (more precisely \ii{Lie renormalized
forms} \emph{Lie renormalized forms} \cite{GaeLRF}); we refer to
\cite{Gsimp} for this. $SU(2)$-related dynamics is also studied in
\cite{GaeRod17}. \EOR

\section{Finite normal forms}
\label{sec:finite}

As discussed above, for systems enjoying an external, ``physical''
symmetry $G$, the normal form corresponds to polynomial vectors
which are symmetric under \emph{both} $G$ and the symmetry $G_0$
identified by the linear part of the system itself (it may happen
that $G_0 \subseteq G$). This condition can, in come cases, be
quite restrictive, and in particular it can happen that there is
only a finite dimensional linear space of vectors satisfying it.
In this case, we have a \ii{finite normal form} \emph{finite
normal form}, i.e. the most general normal form will have a finite
number of terms.

We note that this can be enforced already by the $G_0$ symmetry
alone, as for systems whose linear part satisfies the Poincar\'e
condition (see Appendix \ref{sec:conv}), but here we discuss --
following \cite{GWfinite} -- cases where it is the interplay of
$G_0$ and $G$ to produce this effect.

We denote by $M$ the algebra of ($n \times n$, real) matrices in
$G_0 \oplus G$; to this is associated an algebra $\mathcal{M}$ of
linear vector fields: to any $B \in M$ is associated the vector
field $X_B = (B x) \nabla$.  We are thus interested in the
centralizer $C(\mathcal{M})$ of this algebra in the set of vector
fields. Consider a scalar polynomial function $\Phi : R^n \to R$.
This is a (polynomial) \ii{relative invariant} \emph{relative
invariant} of $\mathcal{M}$ if for all $B \in M$ it results \beq
X_B (\Phi ) \ = \ \mu (B) \ \Phi \ ; \eeq conversely, the set of
functions for which this holds with a given function $\mu : M \to
R$ is denoted as $I_\mu (\mathcal{M})$. Obviously $I_0
(\mathcal{M} )$ corresponds to (polynomial) usual, or \ii{absolute
invariants} \emph{absolute}, invariants.

It is easy to prove \cite{GWfinite} that:

\medskip\noindent
{\bf Proposition 3.} {\it If $I_0 (\mathcal{M})$ is not the full
algebra of polynomials in $R^n$, then $C(\mathcal{M})$ has
infinite dimension. If $C(\mathcal{M})$ has infinite dimension,
then some $I_\mu (\mathcal{M})$ has infinite dimension.}

\medskip\noindent
{\bf Proposition 4.} {\it If $C(\mathcal{M})$ is
infinite-dimensional, then $\mathcal{M}$ admits nontrivial
rational invariants. If moreover either $[\mathcal{M},\mathcal{M}]
= \mathcal{M}$ or $\mathcal{M}$ is solvable, then $I_0
(\mathcal{M})$ is nontrivial.}

\medskip\noindent
{\bf Proposition 5.} {\it Let $\mathcal{M}$ be such that $I_0
(\mathcal{M} ) \not= R$ and $\mathcal{L}$ be the Lie algebra of a
compact linear Lie group such that $[\mathcal{L} , \mathcal{M}]
\subseteq \mathcal{M}$. Then $I_0 (\mathcal{M} + \mathcal{L} )$ is
nontrivial.}
\bigskip

These results can be used to characterize situations in which
$C(\mathcal{M})$ fails to be infinite dimensional; see
\cite{GWfinite} for applications.

\section{Gradient property}
\label{sec:grad}

We say that a group representation has the \ii{gradient property}
\emph{gradient property} \cite{GR} if all the (polynomial) vector
functions which are covariant w.r.t. it can be expressed as
gradients -- ordinary or generalized, i.e. symplectic w.r.t. some
symplectic structure -- of invariant (polynomial) scalar
functions.\footnote{We also speak of gradient property \emph{at
order $N$} if this holds for all polynomial covariant vector
function of degree up to $N$.}

This means that albeit the system has not by itself a variational
nature, it can nevertheless be analyzed with the tools of
variational analysis, with an obvious advantage.

It may happen that a full system is not variational and its
symmetry does not has gradient property, but the reduced equations
corresponding to its normal form near a critical point (or a
bifurcation equation describing the change of stability of this)
have the gradient property.

There is a simple way to ascertain if a group representation has
this property. In fact, as well known, the number of polynomial
invariants $s_k$ and covariants $v_k$ of any degree $k$ can be
computed in terms of the (power expansion of the) \ii{Molien
function} \emph{Molien function}.

If $T = T(g)$ is a representation of the Lie group $G$, and
$(T^{\otimes n})_s$ its symmetrized $n$-fold tensor product, the
number of invariants is given by the coefficient $c^0_n$ in the
Molien series \beq \sum_n c^0_n \, z^n \ = \ \frac{1}{|G|} \
\int_G \mathrm{det} \left[ \frac{1}{(1 - z T (g) )} \right] \ d
\nu (g) \ , \eeq where $d \nu (g)$ is the Haar measure on $G$; the
function $1/(1 - z T (g) )$ is called the \emph{Molien function}.

Similarly, the number of covariants is given by the coefficient
$c^1_n$ in the series \beq c^1_n \ z^n \ = \ \frac{1}{|G|} \
\int_G \mathrm{det} \left[ \frac{1}{(1 - z T (g) )} \right] \
\overline{\chi^1 (g)} \ d \nu (g) \ , \eeq where $\chi (g)$ is the
character of $g$ in the $T$ representation.

\medskip\noindent
{\bf Remark 14.} More generally, $$ |G|^{-1} \ \int_G \mathrm{det}
\left[ \frac{1}{(1 - z T^\mu (g) )} \right] \
\overline{\chi^\sigma (g)} \ d \nu (g) $$ measures the
multiplicity of the representation $T^\sigma$ in
$[(T^\mu)^{\otimes n}]_s$; see e.g. \cite{Sat1,Sat2,SatWea}. \EOR
\bigskip

We denote by $s$ the number of \emph{linear} covariants for $T$;
it is then clear that the number $\gamma_n$ of covariants of order
$n$ which can be obtained as generalized gradients of invariant
functions -- i.e. as $\psi = K \nabla \Phi$ with $K$ a matrix and
$\Phi$ an invariant of degree $n+1$ -- is just $\gamma_n  =  s
\cdot c^0_{n+1}$; in general we have $\gamma_n \le c^1_n$, and the
gradient property (at order $N$) is equivalent to having \beq
c^1_n \ = \ s \cdot c^0_{n+1} \eeq at all orders (for all orders
$n \le N$) \cite{GR}.

It turns out that gradient property holds at all orders for the
defining representation of all the $SO(n)$ and $SU(n)$ groups (and
more generally whenever you have a transitive representation). The
results is surely not extendible beyond the defining
representation, as already for $SO(3)$ it is known not to hold at
order $N=4$ for other representations; on the other hand, it holds
up to order $N=3$ for all representations \cite{G85}.

\section{Spontaneous linearization}
\label{sec:linear}

We speak of \ii{spontaneous linearization} \emph{spontaneous
linearization} when the dynamics of a system evolves towards an
asymptotic regime governed by linear (autonomous or
non-autonomous) equations, and this for whatever initial
conditions or at least for whatever initial conditions in a
certain range (possibly, all those not leading to unbounded
solutions).

It happens that this kind of behavior can be guaranteed on the
basis of symmetry considerations alone, i.e. can be present for
\emph{all} dynamical systems with certain symmetry properties.

In particular, consider the case where the group representation is
transitive on the unit sphere $S^{n-1}$ of the carrier space
$R^n$. In this case we have only one polynomial invariant, which
is just $\rho = x_1^2 + ... + x_n^2 $. According to our discussion
in Section \ref{sec:MicNF}, the evolution of $\rho$ is hence
governed by a function of $\rho$ itself alone, \beq \dot{\rho} \ =
\ h (\rho) \ . \eeq As we have only one variable, either the
solutions $\rho (t)$ diverge or they reach some fixed point
$$ \rho^*_k \ = \ \lim_{t \to \infty} \rho (t) \ ; $$
note that there can be different limit points for different
initial conditions $\rho (0)$.

Recalling now Lemma 2, in this case the most general system in
normal form will be \beq \dot{x} \ = \ \sum_{j=0}^s \mu_j (\rho )
\ M_j x \ . \eeq It is obvious that if we look at the asymptotic
behavior for $t \to \infty$, for all initial data such that the
solution does not diverge (and we recall that the normal form is
relevant to the actual dynamics only in a neighborhood of the
origin), this is given by an equation of the form \beq \dot{x} \ =
\ \sum_{j=0}^s \mu_j (\rho^*_k ) \ M_j x \ = \ \sum_{j=0}^s
\mu^*_{jk} \ M_j x \ ,  \eeq which is indeed linear.

\section{Discussion and conclusions}
\label{sec:discussion}

After introducing the basic ideas -- going back to Poincar\'e --
in Normal Forms theory, we have considered several geometric
aspects of vector fields in normal form. In particular, we have
considered how these have built-in symmetry properties, the
relevant group $G_0$ being associated to the linear part of the
vector field itself, and how they can be reduced to the
$G_0$-orbit space. This in turn means that we can introduce new
auxiliary variables associated to the basic invariants for the
$G_0$ action, and their evolution will depend only on the
invariants themselves.

The same approach can be used in connection with resonant terms of
the vector field -- which in view of the normal form construction
do represent \emph{all} the nonlinear terms -- and in this way we
are led to introduce two sets of auxiliary variables, one
associated to sporadic resonances and one to invariance relations.

The relevant point is that the evolution of the enlarged set of
variables $(x,w,\phi)$ is governed by the equations
\eqref{eq:redredufform}, which are \emph{linear} for the $x$ and
$w$. That is, \emph{if} we are able to solve -- or at least to
determine the asymptotic form of solutions -- the autonomous
system of the equation governing the $\phi$ evolution (this the
system in orbit space), then we are left with a set of --
generally non autonomous -- \emph{linear} equations for the $x$
and $w$. A number of examples, showing this approach can be
implemented in practice, have been presented in Section
\ref{sec:MicNF}, which is the hearth of this work.

We have then considered the case where the system has an
``external'', in particular a physical, symmetry $G$. In this case
the considerations presented above can be extended to consider
$G_0 \times G$, and a reduction of normal forms follows. We have
considered in particular the case where $G$ corresponds to one of
the  \emph{classical groups}. We have also remarked that special
situations -- enforced by Symmetry alone -- can be present; in
particular we have briefly considered the case where normal forms
are necessarily \emph{finite}, that in which they enjoy the
\emph{gradient property}, and that where we get \emph{spontaneous
linearization}.

Having discussed the behavior of vector fields in normal form, we
have noted that the correspondence between the original system and
that in normal form is in general an actual -- and not just formal
-- one, only in a small neighborhood of the origin, if any. The
properties of convergence of the series defining the normalizing
transformation can be checked at hand in any concrete application
and for any \emph{finite order} (the full series being in general
not convergent, and at best \emph{asymptotic}); but nevertheless
one would like to have some general notions of, and results about,
convergence available with no need to actually perform the
detailed computation (that is, available \emph{before} embarking
in a generally complex concrete computation). We have discussed
this, providing some general results -- and focusing on those
based on the symmetry properties of the vector field -- in
Appendix \ref{sec:conv}, thus setting all the previous discussion
on a firmer theoretical basis.

We would now like to mention some topics which, due to limited
space (and time), have not been included in the present discussion
but which would be of interest in this context. Here we can only
give brief hints at them, with a few references.

\begin{enumerate}

\item First of all, as stressed in the Introduction, we have not
considered the specific features of Hamiltonian vector fields. In
this case one can deal directly with the Hamiltonian (a single
scalar function) rather than with the vector field ($2 n$
coefficients).

\item We have also not considered ways to make more efficient the
normalization steps. In particular, in the ``standard'' procedure
(the one described here) one has to invert certain operators,
which is a serious hassle in concrete computations and can present
convergence problems. Both problems can be circumvented by
considering \ii{Lie series} \emph{Lie series}, i.e.
transformations corresponding to the finite-time (e.g. $t=1$) flow
of a vector field. Beside the computational advantages of this
approach, in the present context it has to be noted that it
introduces a further geometrization of the whole procedure. The
approach via Lie series is discussed e.g. in \cite{Bro,CGbook};
see also \cite{Fasso} for the Hamiltonian case.

\item As briefly mentioned in Appendix \ref{sec:NFcomp} (see
footnote \ref{foo:renorm}), one can to some extent control the
effect of normalization on higher order terms, and attempt to
obtain a \ii{further normalization} ``further normalization''. In
the most complete (and hence generally only formal) outcome, this
will make that all terms commute among themselves. The theory is
connected with both geometrical and Lie algebraic aspects, and
introduces further constants of motion (existing in more and more
restricted neighborhood of the origin). For further detail see
e.g. \cite{CGbook} and references therein.

\item We have worked with fixed vector fields \eqref{eq:DS}; but
in many cases of physical interest one is interested in vector
fields depending on (one or more) external parameters. One often
wishes to study situations in which as the parameters are varied
the considered fixed point undergoes a \ii{bifurcation}
\emph{bifurcation} \cite{Crawf,CrawfKno, GaePRbif, Gle,GucHol,
IoossA,IoossJ,Ver}. In this case the basic assumption of the
normalization approach -- i.e. non-degeneration of the linear part
-- fails, and one has to consider a generalization of the theory.
We cannot consider this here, but we mention that in this case too
the presence of symmetries introduces several interesting (and
helpful for practical purposes) aspects \cite{Sat1,Sat2,SatWea}.

\item We would also like to mention that the normal forms approach
was initially devised (and can be used in the non-resonant case)
to obtain a \ii{perturbative linearization} \emph{perturbative
linearization} of the system around a given fixed point. It is
natural to wonder -- in particular in cases where this can be
reached -- if one could operate to obtain the same result outside
perturbation theory, i.e. to investigate \ii{non-perturbative
linearization} \emph{non-perturbative linearization}. For a
normalization-related approach to this problem, see e.g.
\cite{BCGM,GM96}.

\item The normal forms approach was of course created in the
context of Classical Mechanics. It is natural to wonder if this
approach does also extend -- and is effective -- also in the
Quantum Mechanics realm. The answer is positive, and in a way
surprisingly so. Note that in this context one could consider
\ii{quantum normal forms} ``quantum normal forms''
\cite{Ali,Eckhardt,Hage,Little}, or quantize classical normal
forms. In this context, one would expect correspondence with
experimental data to be limited to a small neighborhood of the
origin; in the case of an atom or a molecule this means the
fundamental and maybe some of the first excited levels. It is very
remarkable that instead the normalization approach provides very
good quantitative results up to near the ionization or
dissociation level \cite{Buyuk,Joy,Sugny}.

\item Finally, as we have briefly mentioned above, the normal form
approach has been extended to evolution PDEs \cite{ColEck,Eck,
Niko}, including Hamiltonian PDEs \cite{Cra,Gre,Kuk1,Kuk2}. As far
as I know, the geometric approach sketched here has never been
considered in this context.

\end{enumerate}



\section*{APPENDICES}

{\small

\begin{appendix}

\section{The normal forms construction}
\label{sec:NFcomp}

The construction of Normal Forms is a classical topic, but
nevertheless we will show the basic computation leading to
establishing them \cite{ArnGMDE,Elp} (see also
\cite{ArnODE,ArnEMS}), as this is simple, compact, and at the
basis of our discussion. On the other hand, as already stressed in
the Introduction, we will not deal specifically with the
Hamiltonian case, limiting to discuss things at the level of
vector fields. Moreover, we will just consider explicit
computations in the case where {\tt (iii)} in Section \ref{sec:NF}
above is satisfied.

\subsection{Normalization}

The key idea is to consider \emph{near-identity} changes of
coordinates, and to proceed sequentially normalizing  terms of
order two, three, etc. in the natural order.

Thus, let us suppose the system has been normalized up to
$f_{m-1}$, and let us see how (and in which sense) the term $f_m$
can be simplified. We will consider a change of coordinates of the
form \beq \label{eq:Pmap} x^i \ = \ y^i \ + \ h^i_m (y) \ , \eeq
with $h_m$ homogeneous of degree $m+1$ in the $x$.

All we have to do is to insert this change of coordinates into
\eqref{eq:DSx}, keeping track of what happens at order $m+1$; we
will happily loose track of the effects at higher
order.\footnote{Obviously we can loose track of these for the sake
of theoretical considerations, but in actual applications we will
need to keep carefully track of them! See also footnote
\ref{foo:renorm} in this respect.}

As for the l.h.s. of \eqref{eq:DSx}, we just have
$$ {\dot x}^i \ = \ {\dot y}^i \ + \ \left( \frac{\pa h^i_m}{\pa y^j} \right) \,
{\dot y}^j \ := \ B^i_{\ j} \ {\dot y}^j \ , $$ where we have of
course defined $B = ( I  + \pa h_m/\pa y )$.

The computations referring to the r.h.s. of \eqref{eq:DSx} are
also elementary:
\begin{eqnarray*}
f^i_k (x) &=& f^i_k (y + h_m (y)) \ = \ f^i_k (y) \ + \ \left(
\frac{\pa f^i_k}{\pa y^j} \right) \ h_m^j (y) \ + \
\mathtt{h.o.t.} \ .
\end{eqnarray*} Note that $\pa f_k / \pa y$ is of order $k$, hence
the term we have written explicitly is of order $m+k+1$; similarly
the higher order terms would start with a term $(\pa^2 f_k / \pa
y^p \pa y^q) h^p_m h^q_m$ of order $[(k+1 - 2) + (m+1) + (m +1)] =
(2 m + k + 1)$. Recall also that $m \ge 1$.

Thus, up to h.o.t., eq. \eqref{eq:DSx} reads in the new
coordinates as \beq {\dot y}^i \ = \ (B^{-1} )^i_j \ \left[ \sum_k
f_k^j (y) \ + \ \sum_k \left( \frac{\pa f^j_k}{\pa y^\ell} \right)
\ h_m^\ell (y) \right] \ . \eeq To have this in explicit form, we
only have to note that
$$ B^{-1} \ = \ I \ - \ (\pa h_m / \pa y) \ + \ \mathtt{h.o.t.} \ ,
$$ where now the h.o.t. are of order $2 m$ (if $B = I + \b$, then
$B^{-1} = I - \b + \b^2/2 - ...$). Note also that the term $(\pa
f^j_k / \pa y^\ell) h_m^\ell$ is of order $m+k+1$; so only the
term with $k=0$ is relevant (as we work at order $m+1$).

Thus, in the end, keeping only terms up to order $m+1$, we get
\beq \label{eq:NF} {\dot y}^i \ = \ A^i_{\ j} \, y^j \ + \
\sum_{k=1}^m f^i_k (y) \ + \ A^i_{\ j} h^j_m (y) \ - \ \left(
\frac{\pa h^i_m}{\pa y^j} \right) \, A^j_{\ \ell} \, y^\ell \ + \
\mathtt{h.o.t.} \ . \eeq

In other words, all terms $f_k$ with $k < m$ are unchanged, the
term $f_m$ changes according to \beq \label{eq:fm} f_m \ \to \
\widetilde{f}_m \ = \ f_m \ + \ A^i_{\ j} h^j_m (y) \ - \ \left(
\frac{\pa h^i_m}{\pa y^j} \right) \, A^j_{\ \ell} \, y^\ell \ ,
\eeq and higher order term change in a way we are not explicitly
describing here.\footnote{Keeping control on these can of course
be quite interesting, and leads to a further reduction of the
normal form (one also speaks of \ii{further normalization}
``further normalization''); see e.g. \cite{Bai,BaiChu,ForMur} and
the discussion in \cite{GaePRF,GaeLRF}.\label{foo:renorm}}

\subsection{The homological operator}

We define the linear operator \beq \label{eq:L} \mathcal{L} \ := \
\left( A^j_{\ \ell} \, y^\ell \right) \ \frac{\pa }{\pa y^j}  \ -
\ A \ = \ A x \cdot \nabla \ - \ A \ ; \eeq with this,
\eqref{eq:fm} reads \beq \label{eq:hom} f_m \ \to \
\widetilde{f}_m \ = \ f_m \ - \ \mathcal{L} (h_m ) \ . \eeq This
operator $\mathcal{L}$ is also known as the \ii{homological
operator} \emph{homological operator} associated to $A$.

It is quite obvious that:
\begin{enumerate}

\item denoting by $V_m$ the set of vector functions homogeneous of
order $m+1$, $\mathcal{L} : V_m \to V_m$; thus we can consider the
restriction $\mathcal{L}_m$ of $\mathcal{L}$ to $V_m$, and work at
each order with the  finite dimensional linear operators (that is,
matrices) $\mathcal{L}_m$;

\item terms $\delta h_m$ in ${\rm Ker} (\mathcal{L}_m )$ have no
effect whatsoever, at least at order $m$;

\item by a suitable choice of $h_m$ we can eliminate all terms
$f_m$ in ${\rm Ran} (\mathcal{L}_m )$;

\item hence we can reduce to have (up to a finite but arbitrary
order $N$) \emph{only} nonlinear terms $f_m$ in a space
complementary to ${\rm Ran} (\mathcal{L}_m )$, in which case we
say that the system is in \ii{normal form} \emph{normal form} (to
order $N$);

\item the ``suitable choice'' mentioned above is given by $h_m =
\mathcal{L}_m^* (\pi_m f_m)$, where $\pi_m$ is the projector on
${\rm Ran} (\mathcal{L}_m)$, and $\mathcal{L}_m^*$ is the
pseudo-inverse to $\mathcal{L}_m$; the $h_m$ thus determined is
not unique, being defined up to an element in ${\rm Ker}
(\mathcal{L}_m )$.
\end{enumerate}

\medskip\noindent {\bf Remark \ref{sec:NFcomp}.1.} It is maybe also worth stressing
that the transformation at order $m$ will in general produce new
terms of all orders $\ell > m$. Thus even if we start with only
one nonlinear term, the normalization procedure will produce (even
just at first step) nonlinear terms of all higher orders. Some of
these can be eliminated at later stages, but there can be also
resonant terms (see below) which cannot be eliminated by
normalization at their order. \EOR
\bigskip

In the case where $A_n = 0$, i.e. $A = A_s$, the operator
$\mathcal{L}$ is specially simple. To see this, it is convenient
to write $h_m$ as \beq h_m (y) \ = \ \sum_{i=1}^n \sum_{|J|=m+1}
c_{j_1 ... j_n}^i y_1^{j_1} ... y_n^{j_n} \ {\bf e}_i \ := \
\sum_{i=1}^n \sum_{|J|=m+1} c_{J}^i Y^J \ e_i \ , \eeq where ${\bf
e}_i$ is the $i$-th basis vector in ${\bf R}^n$, $|J| =
\sum_{i=1}^n j_i $, the $j_i$ are non-negative integers, and the
sum over $J = (j_1,...,j_n)$ is on all the $J$ satisfying
$|J|=m+1$. Then, with a compact but intuitive notation,
$$ \frac{\pa h^i_m}{\pa y^\ell} \ = \ \sum_{i=1}^n \sum_{|J|=m+1}
j_\ell \ c_{J}^i Y^{J - e_\ell} \ {\bf e}_i \ . $$ On the other
hand, in this case
$$ A y \ = \ \sum_{i=1}^n \lambda_i y^i \ e_i \ . $$ Therefore, in
the end \beq \mathcal{L} (Y^J {\bf e}_i) \ = \ \left( \sum_\ell
\lambda_\ell j_\ell \ - \ \lambda_i \right) \ {\bf e}_i \ . \eeq

It is hence clear that ${\rm Ker} (\mathcal{L}_k )$ is made of the
vectors whose $i$-th component has a monomial $Y^J$ with
$\lambda_\ell j_\ell = \lambda_i$, and conversely that by choosing
suitably the coefficients $c^i_J$ we can generate all terms except
those in the kernel.

\subsection{Scalar product; adjoint homological operator}

It would be convenient to have a notion of orthogonality in the
spaces of vector functions. The natural scalar product in $V_k$
\cite{ArnGMDE} is defined as follows. We take a basis $\xi_{\mu,i}
= Y^\mu {\bf e}_i$ (with $\mu$ a multi-index, $|\mu| = k$) in each
of the spaces $V_k$, and define \beq \label{eq:Asp} \left(
\xi_{\mu,i} , \xi_{\nu,\ell} \right) \ = \ \delta_{\mu,\nu} \
\delta_{i \ell} \ . \eeq

It would be even better if we could choose a scalar product such
that the range and the kernel of $\mathcal{L}_k$ are orthogonal.
This is possible by choosing the \ii{Bargman scalar product}
Bargman scalar product \cite{Bar,Elp}. Denoting \beq \langle y^\mu
, y^{\nu} {\rm Ran}gle \ = \ \pa_\mu y^{\nu}  \ := \
\frac{\pa^k}{\pa y_1^{\mu_1} ... \pa y_n^{\mu_n} } \
y_1^{\nu_1}... y_n^{\nu_n} \ , \eeq and $\mu ! = (\mu_1 !) ...
(\mu_n !)$, we define the scalar product as \beq \label{eq:Bsp}
\left( \xi_{\mu,i} , \xi_{\nu , \ell} \right) \ := \ \delta_{\mu ,
\nu} \ \delta_{i , \ell } \ \mu! \ . \eeq

The real advantage of this scalar product is embodied in the
following \ii{homological operator}

\medskip\noindent
{\bf Proposition 1.} {\it If $\mathcal{L}$ is the homological
operator associated with the matrix $A$, then its adjoint
$\mathcal{L}^+$ w.r.t. the scalar product \eqref{eq:Bsp} is the
homological operator associated to the matrix $A^+$.}
\bigskip

Now there is a natural choice for the space complementary to ${\rm
Ran} ( \mathcal{L}_k )$, namely $[{\rm Ran} (\mathcal{L}_k ) ]^c \
= \ [{\rm Ran} (\mathcal{L}_k ) ]^+$; moreover -- with the choice
\eqref{eq:Bsp} for the scalar product, we have that \beq
\label{eq:ranker} \left[ {\rm Ran} (\mathcal{L}_k ) \right]^+ \ =
\ {\rm Ker} \left[ \mathcal{L}^+ \right] \ . \eeq


\section{Examples of unfolding}
\label{sec:examples}

In this Appendix we briefly illustrate the construction of Section
\ref{sec:MicNF} by some examples (the last one will actually
concern the discussion of Section \ref{sec:symmetry}). Here it
will always be meant that we consider generic perturbations to a
given linear part, and systems in normal form with respect to (the
semisimple part of) this linear part. We will of course use
coordinates adapted to the linear part (that is, eigencoordinates
for the matrix $A$), and as we work in small dimension we denote
these as $x,y,z$ rather than $x_1,x_2,x_3$, for ease of notation.

\subsubsection*{Example 1.} For $k > 1$ a positive integer and
$$ A \ = \ \pmatrix{1 & 0 \cr 0 & k \cr} \ , $$ the only resonant
vector is $v = (0,x^k)$ (this is associated to a sporadic
resonance), thus the most general system in normal form reads
\begin{eqnarray*}
\dot{x} &=& x \\
\dot{y} &=& k \, y \ + \ c \, x^k \ , \end{eqnarray*} with $c$ an
arbitrary real constant.  Introducing $w \simeq x^k$, the system
unfolding reads
\begin{eqnarray*}
\dot{x} &=& x \\
\dot{y} &=& k \, y \ + \ c \, w \\
\dot{w} &=& k \, w \ . \end{eqnarray*} Note that the manifold $M$
identified by $\psi := w - x^k = 0$ is obviously invariant under
this flow; in fact,
$$\frac{d \psi}{d t} \ = \ \dot{w} \ - \ k \, x^{k-1} \, \dot{x} \ = \
k \, w \ - \ k \, x^k \ = \ k \ \psi \ . $$ The solution to this
three-dimensional system is immediately obtained,
$$ x(t) = x_0 \, e^t \ , \ \ y(t) = y_0 \, e^{k t} \ + \ (c k w_0)
\, t \, e^{k t} \ , \ \ w(t) = w_0 \, e^{k t} \ ; $$ restricting
this to the manifold $M$ and projecting to the two-dimensional
space spanned by $x$ and $y$, we get
$$ x(t) \ = \ x_0 \, e^t \ , \ \ y(t) \ = \ \left[ y_0 + (c_1 k x_0 )
t \right] \, e^{k t} \ . $$

\subsubsection*{Example 2.} Let us consider (the perturbation of)
a system of two oscillators with irrational frequencies, i.e.
$$ A \ = \ \pmatrix{0 & -1 & 0 & 0 \cr 1 & 0 & 0 & 0 \cr 0 & 0 & 0 & - \omega \cr
0 & 0 & \omega & 0 \cr} $$ with $\omega$ real and irrational. Now
we have eigenvalues $(\lambda_1,\lambda_2 , \lambda_3 , \lambda_4)
= (- i , + i , - i \omega ,+i \omega)$. There are no sporadic
resonances apart from the trivial ones of order one, and two
invariance relations, given obviously by $\lambda_1 + \lambda_2 =
0$, $\lambda_3 +\lambda_4 = 0$; correspondingly we have two
invariant functions, $\rho_1 = x_1^2 +x_2^2$, $\rho_2 = x_3^2 +
x_4^2$. Thus the general normal form for perturbations of this
system is written, in vector notation, as
$$ \dot{\xi} \ = \ M \ \xi \ , $$
where $\xi = ( x_1,x_2,x_3,x_4)$ and $M$ is a $4 \times 4$ matrix
of the form
$$ M \ = \ \pmatrix{ \a & - \b & 0 & 0 \cr \b & \a & 0 & 0 \cr
0 & 0 & \gamma & - \eta \cr 0 & 0 & \eta & \gamma \cr} $$ with
$\a, \b,\gamma , \eta$ polynomial functions of $\rho_1 , \rho_2$.

The evolution of these is given by
\begin{eqnarray}
\dot{\rho}_1 &=& 2 \ \a (\rho_1 , \rho_2) \ \rho_1 \ , \nonumber \\
\dot{\rho}_2 &=& 2 \ \gamma (\rho_1 , \rho_2) \ \rho_2 \ ;
\label{eq:systex2} \end{eqnarray} if we are able to solve this
system, then we write $ \a (t)  := \a [ \rho_1 (t) , \rho_2 (t) ]$
and the like, and we are reduced to studying a linear
(time-dependent) four-dimensional system, $\dot{\xi} = M(t) \xi $
(which actually decouples into two two-dimensional ones).

Note that even if we are not able to solve the above system
\eqref{eq:systex2}, its nature could (depending on the functions
$\sigma$ and $\gamma$) allow us, via the Poincar\'e-Bendixson
theorem, to be sure it will go asymptotically either to a constant
or to a periodic motion. Correspondingly, the four-dimensional
linear time-dependent system to be solved would become
(asymptotically) a time-independent or time-periodic one.

\medskip\noindent
{\bf Remark \ref{sec:examples}.1.} The linear part of this system
generates the irrational flow on the torus ${\bf T}^2$, so the
(compact!) closure of this group is the full ${\bf T}^2$;
correspondingly we have invariants associated to the full ${\bf
T}^2$ action. \EOR

\subsubsection*{Example 3.} Consider now two oscillators in 1:1 resonance, i.e.
$$ A \ = \ \pmatrix{0 & -1 & 0 & 0 \cr 1 & 0 & 0 & 0 \cr 0 & 0 & 0 & - 1 \cr
0 & 0 & 1 & 0 \cr} \ . $$ The eigenvalues are
$(\lambda_1,\lambda_2 , \lambda_3 , \lambda_4) = (- i , + i , - i
,+i )$. There are no sporadic resonances apart from the trivial
ones of order one, but four invariance relations, given obviously
by $\lambda_1 + \lambda_2 = 0$, $\lambda_3 +\lambda_4 = 0$,
$\lambda_1 + \lambda_4 = 0$, $\lambda_2 + \lambda_3 = 0$; note
that these are linearly dependent, but none of this can be written
polynomially in terms of the others. Correspondingly we have four
(algebraically but not functionally independent) invariant
functions, which can be chosen in a non-unique way; e.g. we choose
$\rho_1 = x_1^2 +x_2^2$, $\rho_2 = x_3^2 + x_4^2$, $\rho_3 = x_1
x_3 + x_2 x_4$, $\rho_4= x_1 x_4 - x_2 x_3$. The centralizer of
$A$ is now an eight-dimensional algebra; in terms of the
two-dimensional matrices
$$ I \ = \ \pmatrix{1 & 0 \cr 0 & 1 \cr} \ , \ \
J \ = \ \pmatrix{0 & - 1 \cr 1 & 0 \cr} \ , $$ its generators can
be written in block notation as
\begin{eqnarray*} B_1 = \pmatrix{I & 0 \cr 0 & 0 \cr} &,& \
B_2 = \pmatrix{0 & 0 \cr 0 & I \cr} \ , \ B_3 = \pmatrix{0 & I \cr
I & 0 \cr} \ , \
B_4 = \pmatrix{0 & J \cr -J & 0 \cr} \ ; \\
S_1 = \pmatrix{J & 0 \cr 0 & 0 \cr} &,& \ S_2 = \pmatrix{0 & 0 \cr
0 & J \cr} \ , \ S_3 = \pmatrix{0 & -I \cr I & 0 \cr} \ , \ S_4 =
\pmatrix{0 & J \cr J & 0 \cr} \ . \end{eqnarray*} Note that
$B_i^+= B_i$, $S_i^+= - S_i$; actually our choice of the $\rho_i$
corresponds to $\rho_i = ( \xi , B_i \xi )$, where $(.,.)$ is the
standard scalar product in $R^4$.

If now we consider (with the same notation as in the previous
example) perturbations of the linear system $\dot{\xi} = A \xi$,
the normal form for these  will be written as $\dot{\xi} = M
(\rho) \xi$, where
$$ M \ = \ \pmatrix{\a & - \b & \gamma & - \eta \cr \b & \a & \eta & - \gamma \cr
\mu & - \nu & \sigma & - \tau \cr \nu & \mu & \tau & \sigma \cr} \
= \ \pmatrix{\a I + \b J & \gamma I + \eta J \cr \mu I + \nu J &
\sigma I + \tau J \cr} \ , $$ and the $\a , ... , \tau$ are
polynomial functions of $\rho_1 , ... , \rho_4$.

Note that while the linear system is invariant under the exchange
of the two oscillators, or under a simultaneous rotation in the
$(x_1 x_3)$ and the $(x_2 x_4)$ planes, the general normal form
does not enjoy any property of this type.

The evolution of the invariant functions is compactly written as
\beq \label{eq:systex3} \dot{\rho}_a \ = \ \left( \xi \, , \, (B_a
M + M^+ B_a ) \, \xi \right) \ ; \eeq as $M$ is in general a
nonlinear function of the $\rho$, this is a system of four
nonlinear equations and in general we are not able to solve them,
nor we can use the \ii{Poincar\'e-Bendixson theorem}
Poincar\'e-Bendixson theorem as in the previous Example in order
to get information about its asymptotic behavior. If we are able
by some means to determine some (stable) stationary or periodic
solution to \eqref{eq:systex3}, we can then determine the
corresponding solutions to the linear system $\dot{\xi} = M [ \rho
(t)] \xi$.

\subsubsection*{Example 4.} Our last example concerns the situation
discussed in Section \ref{sec:symmetry}. We consider a system
which is the perturbation of two identical oscillators, coupled in
such a way to have an exchange symmetry (that is, the system is
covariant under the simultaneous exchanges $x_1 \leftrightarrow
x_3$, $x_2 \leftrightarrow x_4$). Then the linear part is again
$$ A \ = \ \pmatrix{0 & -1 & 0 & 0 \cr 1 & 0 & 0 & 0 \cr 0 & 0 & 0 & - 1 \cr
0 & 0 & 1 & 0 \cr} $$ as in Example 3, and the same considerations
as in Example 3 above apply, \emph{but} only nonlinear resonant
terms respecting the exchange  symmetry are allowed.

Note that under this exchange, $\rho_3$ is invariant, while
$\rho_1 \leftrightarrow \rho_2$ (thus we should consider as
invariant $\rho_1 + \rho_2$), and $\rho_4 \to - \rho_4$. Thus we
have only two invariants, say $\phi_1 = \rho_1 + \rho_2$ and
$\phi_2 = \rho_3$.

Correspondingly, we should consider the simultaneous centralizer
of $A$ and of the exchange matrix
$$ E \ = \ \pmatrix{0 & 0 & 1 & 0 \cr 0 & 0 & 0 & 1 \cr
1 & 0 & 0 & 0 \cr 0 & 1 & 0 & 0 \cr} \ = \ \pmatrix{ 0 & I \cr I & 0 \cr} \ . $$

This is a four-dimensional algebra, spanned by the matrices
$$ B_1 +B_2 \ , \ B_3 \ ; \ S_1 +S_2 \ , \ S_4 \ . $$
Correspondingly, the most general normal form $\dot{\xi} = M
(\rho) \xi$ is now written in terms of matrices $$ M \ = \
\pmatrix{\a & - \b & \gamma & - \eta \cr \b & \a & \eta & - \gamma
\cr \gamma & - \eta & \a & - \b \cr \eta & \gamma & \b & \a \cr} \
= \ \pmatrix{\a I + \b J & \gamma I + \eta J \cr \gamma I + \eta J
& \a I + \b J \cr} \ , $$ and the $\a , ... , \eta$ are polynomial
functions of $\phi_1 , \rho_2$.

This should be compared with the situations described in Example
3.

\section{Hopf and Hamiltonian Hopf bifurcations}
\label{sec:Hopf}

The approach described in the Section \ref{sec:MicNF} can also be
used to study systems at a \ii{Hopf bifurcation} Hopf or
\ii{Hamiltonian Hopf bifurcation} Hamiltonian Hopf bifurcation.
Note that in these cases, at difference to what happens e.g. in a
\ii{pitchfork bifurcation} pitchfork bifurcation, the linear part
of the system at the bifurcation does not vanish; this allows to
apply the normal form approach (and hence also our method) also at
the bifurcation.

\subsection{Hopf bifurcation}

In the case of Hopf bifurcation \cite{Crawf,CrawfKno, GaePRbif,
Gle,GucHol, IoossA,IoossJ,Ver}, the linear part of the system at
the bifurcation is described by \beq A \ = \ \pmatrix{ 0 & -
\omega_0 \cr \omega_0 & 0 \cr} \ , \eeq where $\omega_0 \not= 0$
is a real parameter representing the frequency (of the bifurcating
periodic solutions) at the bifurcation.

The eigenvalues are obviously $\lambda_\pm = \pm i \omega_0$. So
we have no sporadic resonances, and one invariance relation,
$\lambda_+ + \lambda_- = 0$; the associated invariant is $\rho =
x^2 + y^2 \ge 0$. The most general system in normal form is hence
\begin{eqnarray*}
\dot{x} &=& \a (\rho , \mu) \ x \ - \ \b (\rho , \mu ) \ y \\
\dot{y} &=& \b (\rho , \mu) \ x \ + \ \a (\rho , \mu ) \ y \ ,
\end{eqnarray*}
where we have allowed the system to also depend on an external
parameter (driving the bifurcation) $\mu$. In this case we get
immediately $$ \dot{\rho} \ = \ 2 \ \rho \ \a (\rho , \mu ) \ . $$

In the standard setting for Hopf bifurcation,
$$ \a (\rho , \mu) \ = \ \mu \, - \, c \, \rho \ ; \ \ \b (\rho ,
\mu) \ = \ \omega_0 \ + \ b (\rho , \mu ) $$ where $b(0,0) = 0$
(here the bifurcation takes place in the origin at $\mu = 0$).

In our approach, writing $\rho = \phi$ to recover our general
notation, we pass to study a three-dimensional system
\begin{eqnarray*}
\dot{x} &=& \mu \, x \ - \ \omega_0 \, y \ - \ c(\phi) \, x \ - \
b(\phi,\mu) \, y \\
\dot{y} &=& \omega_0 \, x \ + \ \mu \, y \ + \ b(\phi,\mu) \, x \
- \
c(\phi) \, y \\
\dot{\phi} &=& 2 \, \a(\phi , \mu) \ \phi \ . \end{eqnarray*} The
invariant $\phi (t)$ can grow indefinitely; but if this is not the
case, it will approach one of the zeros of the function $\a
(\phi,\mu)$, call it $\phi_0$. Then we get asymptotically the
linear system
\begin{eqnarray*}
\dot{x} &=& - \ \left[ \omega_0 \ + \ b (\phi_0 , \mu ) \right] \, y  \\
\dot{y} &=& \left[ \omega_0 \ + \ b(\phi_0 ,\mu) \right] \, x  .
\end{eqnarray*}
The standard analysis of Hopf bifurcation is thus recovered.

\subsection{Hamiltonian Hopf bifurcation}

In the case of Hamiltonian Hopf bifurcation \cite{VdM1,VdM2} the
linear part of the system at the bifurcation is \beq A \ = \
\pmatrix{\mu & - \omega & 0 & 0 \cr \omega & \mu & 0 & 0 \cr 0 & 0
& - \mu & - \omega \cr 0 & 0 & \omega & - \mu \cr} \ = \ \pmatrix{
\mu I + \omega J & 0 \cr 0 & - \mu I + \omega J \cr} \ , \eeq
where $\mu \not= 0$ and $\omega \not= 0$ are real parameters (note
that $\mu = 0$ corresponds to a pair of oscillators in 1:1
resonance; in applications, $\mu$ is the external control
parameter and when it goes through zero we have the bifurcation).

We thus have eigenvalues $\lambda_{\pm \pm} = \pm \mu \pm i
\omega$; for generic $\mu$ there are no sporadic resonances, and
there are two invariance relations, $$ \lambda_{++} \ + \
\lambda_{--} \ = \ 0 \ ; \ \ \lambda_{+-} \ + \ \lambda_{-+} \ = \
0 \ ;
$$ the associated invariants are $\phi_1 = x_1 x_3 + x_2 x_4$ and
$\phi_2 = x_1 x_4 - x_2 x_3$.

It is convenient to introduce the two-dimensional vectors $\eta_1
= (x_1,x_2)$, $\eta_2 = (x_3,x_4)$, $\phi = (\phi1 , \phi_2)$. The
matrices in the centralizer of $A$ are written in terms of real
constants $\a_k$, $\b_k$ as block-diagonal ones, of the form $ M =
\mathtt{diag} (\a_1 I + \b_1 J ,  \a_2 I + \b_2 J )$.
Correspondingly, systems in normal form will be  given by
$$ \dot{\eta} \ = \ \pmatrix{ \a_1 (\phi ) \, I \ + \
\b_1 (\phi ) \, J & 0 \cr 0 & \a_2 (\phi ) \, I \ + \ \b_2 (\phi )
\, J \cr} \ \eta \ . $$ The functions $\a_k , \b_k$ (which can
also depend on $\mu, \omega$, albeit we omitted to write
explicitly this dependence) can be written as
$$ \a_k \ = \ (-1)^{k+1} \, \mu \ + \ a (\phi ) \ , \ \ \b_k \ = \
\omega \ + \ b_k (\phi ) \ ; \ \ a_k (0) \, = \, 0 \, = \, b_k (0)
\ . $$ Note that the system is Hamiltonian, i.e. preserves the
symplectic form $d x^1 \wedge d x^2 + d x^3 \wedge d x^4$, if and
only  if $\a_2 = - \a_1 $ and $\b_2 = \b_1$ (these relations are
always satisfied at the linear level).

The system is hence described, proceeding with our unfolding
procedure, as
\begin{eqnarray*}
\dot{\eta}_1 &=& \left[ \left( \mu + a_1 (\phi) \right) \, I \ + \
\left( \omega - b_1
(\phi) \right) \, J \right] \ \eta_1 \ , \\
\dot{\eta}_2 &=& \left[ \left( - \mu + a_2 (\phi) \right) \, I \ +
\ \left( \omega -
b_2 (\phi) \right) \, J \right] \ \eta_2 \ ; \\
\dot{\phi} &=& \left[ \left( a_1 (\phi) + a_2 (\phi) \right) \, I
\ + \ \left( b_2 (\phi) - b_1 (\phi) \right) \, J \right] \ \phi \
.
\end{eqnarray*}

Note that if $\a_k,\b_k$ are such that the system is Hamiltonian,
$\phi$ is constant and we are always reduced to a linear system on
each level set of $\phi = (\phi_1,\phi_2)$; if the system is not
Hamiltonian we deal however with a two dimensional system and
under the standard conditions for a bifurcation to take place --
i.e. if $|\phi |$ cannot grow indefinitely -- the
\ii{Poincar\'e-Bendixson theorem} Poincar\'e-Bendixson theorem
applies.

At the bifurcation point $\mu = 0$ we are in the framework of
Example \ref{sec:examples}.4 above.

\section{Symmetry and convergence for normal forms}
\label{sec:conv}

We have seen that vector fields in normal forms can be
characterized in terms of a symmetry property. However, symmetry
is independent of the coordinate description. This suffices to
conclude, as first emphasized by J. Moser \cite{Mos1,Mos2}, that a
vector field which lacks symmetry cannot be conjugated to its
normal form (that is, in this case the series describing the
normalization procedure are only formal ones).

It is somewhat surprising that -- modulo certain, not always
trivial, conditions -- the converse is also true. That is, a
suitable symmetry of the vector field suffices, together with
other assumptions, to guarantee the convergence of the normalizing
transformation. This theory has been developed by Bruno,
Markhashov, Walcher and Cicogna
\cite{Bru1,Bru2,BruWal,CicCNF,Mar,Walconv}; see also works by Ito,
Russmann and Vey in the Hamiltonian case \cite{CWaam}, which we
will not touch upon.

The matter is discussed in some length by Cicogna and Walcher
\cite{CWaam}; here we will just report the main results, using
freely the notation established above -- in particular we give for
understood that we are dealing with transformation of a system
\eqref{eq:DS} into normal form near an equilibrium, and that $A =
(Df)(0)$ is the matrix describing the linearization of the system
at the given equilibrium.

First of all, we recall a \ii{Poincar\'e criterion} classical
result.

\medskip\noindent
{\bf Poincar\'e criterion.} {\it Let $\lambda_1,...,\lambda_n$ be
the eigenvalues of the matrix $A$; if the convex hull of these
does not contain zero, then the normalizing transformation is
convergent and the normal form is analytic.}

\medskip\noindent
{\bf Remark \ref{sec:conv}.1.} Note that with this assumption on
the spectrum of $A$ (which are also stated saying that $A$ belongs
to a \ii{Poincar\'e domain} \emph{Poincar\'e domain}), we have no
invariance relations, and only a finite number (possibly zero) of
sporadic resonances. \EOR
\bigskip

We will then introduce two rather general
conditions\footnote{Their names, taken together, have a slight
taste of blasphemy; unfortunately these names are by now well
established in the literature.}, first considered by Bruno
\cite{Bru1,Bru2} (here summation over the dummy index $i$ is
implied, for ease of writing).

\medskip\noindent
{\bf Condition $\a$.} {\it The vector field $f(x)$ with linear
part $A x$ is said to satisfy \emph{condition $\a$} if \beq
\label{eq:cA} f(x) \ = \ [ 1 + \a (x) ] \ A \, x \eeq for some
scalar-valued power series $\a (x)$.}

\medskip\noindent
{\bf Condition $\omega$.} {\it Let $\Lambda = \{ \lambda_1 , ... ,
\lambda_n \}$ be the spectrum of $A = (Df)(0)$. Denote by
$\omega_k$ (with $k > 0$ any integer) the minimum of $ | q^i
\lambda_i \ - \ \lambda_j |$ for all $j=1,..,n$ and all $n$-tuples
of non negative integers $q^i$ such that $q^i \lambda_i \not=
\lambda_j$ and $|q| = q^1 + ... + q^n$ satisfies $1 < |q| < 2^k$.
Then if \beq \label{eq:cOm} \sum_{k=1}^\infty 2^{-k} \ \log \left(
\omega_k^{-1} \right) \ < \ \infty \eeq we say that
\emph{condition $\omega$} is satisfied.}

\medskip\noindent
{\bf Remark \ref{sec:conv}.2.} Condition $\omega$ is a mild one,
guaranteeing there is no accumulation of small denominators. We
will always assume this is satisfied in the forthcoming
discussion. \EOR

\medskip\noindent
{\bf Proposition \ref{sec:conv}.1.} \cite{Bru1}. {\it If $A$
satisfies condition $\a$, and $f$ can be taken  by means of a
sequence of Poincar\'e transformations to a normal form which
satisfies condition $\a$, then there is a convergent normalizing
transformation for $f$.}
\bigskip

We will now consider the symmetric case. In this case one can aim
at normalizing both the dynamical vector field and the symmetry
one; moreover, favorable properties of the latter one can extend
their benefit also on the former one.

\medskip\noindent
{\bf Proposition \ref{sec:conv}.2.} \cite{CWaam} {\it Let the
dynamical system \eqref{eq:DS}, with $f(x) = A x + F (x)$,  admit
an analytic symmetry vector field $Y = g^i (x) \pa_i$, with $g (x)
= B x + G(x)$. Assume that either $(i)$ $B$ belongs to a
Poincar\'e domain, or $(ii)$ there is a coordinate transformation
taking $g$ into a normal form satisfying condition $\a$. Then
there is a convergent normalizing transformation which takes $g$
into normal form and moreover maps $f$ into $\widetilde{f} = A x +
\widetilde{F} (x)$, where $\widetilde{F}$ is resonant with $B$
(not necessarily with $A$).}
\bigskip

We now recalling the statement -- and the notation -- of Lemma 2
above, see Section \ref{sec:symmNF}. As mentioned there, if $s=0$
we are in the case where assumption $\a$ is surely satisfied, and
hence there is a convergent normalizing transformation. The
following theorem considers the more general setting.

\medskip\noindent
{\bf Proposition \ref{sec:conv}.3.} \cite{CicCNF} {\it Let the
normal form for $f$ be written in the form
$$ \widehat{f} (x) \ = \ [1 \, + \, \a (x) ] \, A \, x \ + \
\sum_{j=1}^s \mu_j (x) \, M_j \, x \ := \ [1 \, + \, \a (x) ] \, A
\, x \ + \ \widehat{F} (x) \ , $$ with $\mu_j$ and $M_j$ as in
Lemma 2, $s\not=0$, and $\widehat{F} \not= 0$. Assume that:
\begin{enumerate}
\item $f$ admits an analytic symmetry $g (x) = B x + G(x)$ with $B
= k A$ for some (possibly zero) constant $k$, and $G$ fully
nonlinear; \item if $S = \sum_{j=1}^s \nu_j (x) M_j$ (with $X_0
(\nu) = 0$) nd $(DS)(0)= 0$, the equation $\{\widehat{F} , S \} =
0$ has only trivial solutions $S = c \widehat{F} (x)$.
\end{enumerate}
Then $f$ can be taken to normal form by means of a convergent
normalizing transformation.}
\bigskip

A simple yet relevant and clarifying application of these result
is provided by the problem of the \ii{isochronous center}
\emph{isochronous center}, see the discussion in \cite{CWaam}.

\end{appendix}

}


\let\a\relax
\let\b\relax
\let\pa\relax
\let\eeq\relax
\let\beq\relax
\let\eqref\relax
\let\EOR\relax

\end{document}